\documentstyle[aps, preprint]{revtex}

\begin{document}

\title{Cosmological Perturbations From Inhomogeneous Reheating, Freeze-Out, and Mass Domination}

\author{ Gia Dvali$^a$, Andrei Gruzinov$^{a}$, and Matias Zaldarriaga$^{a,b,c}$}

\address{$^a$ Center for Cosmology and Particle Physics, Department of Physics, New York University, NY 10003}
\address{$^b$ Department of Physics, Harvard University, Cambridge, MA 02138}
\address{$^c$ Department of Astronomy, Harvard University, Cambridge, MA 02138}

\date{April 13, 2003}
\maketitle

\setcounter{footnote}{0} \setcounter{page}{1}
\setcounter{section}{0} \setcounter{subsection}{0}
\setcounter{subsubsection}{0}

\begin{abstract}

We generalize a recently proposed mechanism for the origin of
primordial metric perturbations in inflationary models. Quantum
fluctuations of light scalar fields during inflation give rise to
super-horizon fluctuations of masses and reaction rates of various
particles. Reheating, freeze-out, and matter-domination processes
become inhomogeneous and generate super-horizon metric
perturbations. We also calculate the degree of non-Gaussianity
$f_{nl}$ for this new model of cosmological perturbations. The precise
value of $f_{nl}$ depends on the specific models, but
$|f_{nl}|\sim$few is a natural lower bound for our mechanisms. This is
much larger than the currently assumed theoretical value $f_{nl}\sim
tilt \lesssim 0.05$, and is thought to be observable. In a
particularly attractive model of inhomogeneous mass-domination, the
non-Gaussianity of perturbations generated by our mechanism is simply $f_{nl}=5$, irrespective of the detailed structure of the underlying field theory.

\end{abstract}
\section{Introduction}

During inflation the energy density is dominated by the potential
energy of a slowly rolling scalar field, the inflaton. At the end of
inflation this energy density has to be converted into normal
particles, reheating the universe and starting the standard phase of
the hot big bang. In \cite{dgz} we suggested that if the inflaton
decay rate $\Gamma$ varied in space, density perturbations would be
generated during reheating independently of those generated by the
standard inflationary mechanism. If two different regions of the
universe had different $\Gamma$s then effectively the inflaton would
decay into radiation first in one region and then in the other. During
the time one region is filled with radiation while the other one is
not, the universe expands at a different rate in each region,
resulting in density perturbations when reheating is finished. The
decay rate of the inflaton is determined by the expectation values of
some scalar fields. If those scalar fields were light during inflation
they fluctuated, leading to density perturbations through the proposed
mechanism. Here we extend and generalize this model (\S II).

We also calculate the non-Gaussianity of the metric perturbations
generated by our mechanism. The standard inflationary model predicts
unobservably small non-Gaussianities, while our models generically
predict non-Gaussianities of potentially observable magnitude (\S
III).

\section{Mass-Domination Mechanism}

\subsection{The Mechanism}

We first discuss a generalization of our original mechanism
\cite{dgz}. Assume that density perturbations created during
inflation, as well as during reheating, are negligible.  So right
after reheating the Universe is filled with radiation of uniform
energy density and temperature $T_R$.  Assume that the mass and the
decay rate ($\Gamma$) of some of the created particles (call them
$\psi$) is set by the VEV of a scalar field $\phi$,
\begin{equation}
 M \,  = \, \lambda\, \phi
\label{mass}
\end{equation}
where $\lambda$ is a coupling constant.  We shall assume that the mass
of $\phi$ during inflation is smaller than the Hubble parameter.
Thus, inflationary fluctuations of $\phi$ on superhorizon scales get
imprinted into the mass of $\psi$-quanta.  As long as the temperature
is much larger than $M$, the perturbations in mass do not contribute
into the energy density. However, once the temperature drops below
$M$, the mass-fluctuations become important. For the mass fluctuations
to be imprinted into the density perturbations, it is essential that
$\psi$ dominates (or at least becomes a significant component of) the
energy density of the Universe during some period.  That is, the decay
rate of $\psi$ must satisfy
\begin{equation}
 \Gamma \, < \, H_{D}
\label{hcon}
\end{equation}
where $H_{D}$ the Hubble parameter at the moment when $\psi$ starts
dominating the energy density. For simplicity, we shall assume that
the annihilation rate of $\psi$ is much smaller than
$M^2/M_{Pl}$. Then $\psi$ start dominating as soon as they become
non-relativistic ($T \sim M$, assuming that $\psi$ were thermalized at
some early time). The corresponding Hubble parameter is $H_D \, \sim
\, M^2/M_{Pl}$.

If (\ref{hcon}) is satisfied, the perturbations in $M$ and $\Gamma$
get translated into the density perturbations.  Due to the variation
of $M$ and $\Gamma$, the interval of $\psi$-domination in different
regions of the Universe will be different leading to density
perturbations at the end of the process.

 We shall now derive the magnitude of the resulting density
perturbations.  We compare final radiation energies (after
$\psi$-decay) in different superhorizon regions, for the same value of
the scale factor $a$.  It is simplest to compare energy density in any
given region to the one in radiation, which scales as
\begin{equation}
\label{rad}
\rho_{rad} \, \propto \, a^{-4},
\end{equation}
and is the same in all the regions of interest.  The $\psi$-energy
density in a region with mass $M$ and decay rate $\Gamma$ scales as
radiation for all the values of $a$, except the interval of its
domination, that is, between the domination and the decay. Domination
starts at $a = a_{domination}$ when
\begin{equation}
\rho \, = \, \rho_{domination}= M^4
\end{equation}
and ends at $a = a_{decay}$, when
\begin{equation}
\rho = \rho_{decay}\, = \, \Gamma^2M_{Pl}^2
\end{equation}
Outside this interval $a_{decay} > a > a_{domination}$, the energy
density in all the domains scales as radiation and is independent of
either $M$ or $\Gamma$. However, during the domination interval energy
of $\psi$ scales as matter
\begin{equation}
\label{mat}
\rho \, \propto \, a^{-3}
\end{equation}
and becomes different in different domains.
Thus, we have
\begin{equation}
\left ( {a_{decay} \over a_{domination}} \right )^3 = {\rho_{domination}\over
 \rho_{decay}} = M^4 \Gamma^{-2} M_{Pl}^{-2}
\label{interval}
\end{equation}
The final energy density stored in $\psi$, right before the
decay is 
\begin{equation}
\rho \propto {a_{decay} \over a_{domination}}  \rho_{rad} 
= M^{4/3} \Gamma^{-2/3} M_{Pl}^{-2/3} \rho_{rad}
\label{final}
\end{equation}
The resulting density perturbations are given by
\begin{equation}
{\delta \rho \over\rho} \, =  \, 
{4\over 3}{ \delta M \over M } \, - \, {2 \over 3} {\delta\, \Gamma \over \Gamma}
\label{delta}
\end{equation}

Depending on the underlying model, one can consider several
possibilities.  For instance, for $M = constant$, this recovers our
previous result\cite{dgz}, where now we are interpreting $\Gamma$ as
the decay rate of the inflaton rather than $\psi$.  The results carry
over from one scenario to the other because during reheating, when the
oscillating inflaton dominates the energy density we also have $\rho
\propto a^{-3}$.  Another interesting case for the realistic model
building is $\Gamma \propto M$, for which we get
\begin{equation}
{\delta \rho \over\rho} \, =  \, 
{2\over 3}{ \delta M \over M }
\label{delta1}
\end{equation}

\subsection{Constraints}

 The necessary requirement for our mechanism is that an effective mass
of $\phi$ must remain smaller than the Hubble parameter during
inflation. However, this relation may or may not be violated during
$\psi$-domination period. If it is violated, then $\phi$ will start
oscillations during the $\psi$ domination epoch. The calculation of the
resulting density perturbations for oscillating $\phi$ is more
involved and will not be discussed here.

 The simplest situation is when the effective mass of $\phi$ remains
smaller than the Hubble parameter all the way until the $\psi$
particles decay.  This requirement strongly constraints the initial
value of $\phi$, provided the $\phi$-dependence of $\psi$-mass is
significant.  Below we shall estimate the constraint for the case of
maximal dependence, that is when the entire mass of $\psi$ comes from
$\phi$ according to (\ref{mass}). Generalization for the case of
milder dependence is obvious.

In many cases this puts a non-trivial constraint on the initial value
of $\phi$, as we shall now demonstrate. The crucial point is that at
high temperature the non-zero density of $\psi$-particles generates a
large thermal mass for $\phi$.  For instance, every scalar degree of
freedom that is in thermal equilibrium at temperature $T$, and is
coupled to $\phi$, generates the following contribution to the
$\phi$-mass
 \begin{equation}
m^2_{\phi} \, \sim  \, \lambda^2 T^2   
\label{tm}
\end{equation}
This contribution is there irrespective of $\phi$ itself being in
thermal equilibrium.  This mass may exceed the value of the Hubble
parameter during the $\psi$-domination, unless the following condition
is met
\begin{equation}
\lambda  T_D  \, < \, H_D
\label{tmcondition}
\end{equation}
where $T_D\, = \, \lambda \phi$ is the domination temperature. This implies that $\phi$
must satisfy the following condition. 
\begin{equation}
\,1\, < \, {\phi \over M_{Pl}}
\label{phicondition}
\end{equation}
Notice that (\ref{phicondition}) guaranties that the condition
$m_{\phi} < H$ will be satisfied both before and throughout the
domination. This is easy to understand. Above the domination
temperatures the thermal mass scales as temperature, and quickly
becomes negligible relative to Hubble parameter, which scales as
$T^2$. During the domination period $\psi$-particles are out of
thermal equilibrium and their energy scales as non-relativistic
matter, according to (\ref{mat}). So does the mass of $\phi$,
$m_{\phi}^2 \propto \rho_{\psi}$. Thus, $m_{\phi}$ and $H$ scale in
the same way, and relation $m_{\phi} < H$ is maintained throughout the
domination.

 For typical $\phi$-dependence of $M$ and $\Gamma$, $\phi \sim M_{Pl}$
implies that the amount of density perturbations created by standard
inflationary scenario will not be sub-dominant.  So, our mechanism
will be the primary source of perturbations only in inflationary
scenarios that have no fluctuating inflaton field, e.g., such as
recently proposed ``self-terminated inflation''\cite{self}.  Even in
this case the level of the gravitational wave background should be
significant.

We should again stress that the above constraint in absent in the
models in which $M$ is independent of $\phi$. In these class of models
our source of density perturbations can dominate over the standard
inflationary mechanism by many orders of magnitude.

\subsection{Models}

 We shall now consider some practical implementations
of our mechanism in realistic models. As is obvious from the above,
we are interested in theories in which masses of dominating particles
$M(\phi)$ and/or their decay rate(s) $\Gamma(\phi)$ are functions 
of the fluctuating flat-direction field $\phi$. Since, in general 
$\phi$-dependence of functions $M(\phi), \Gamma(\phi)$ can be very
different, it is hard to talk about
more than an order of magnitude predictions.
However, there is a sub-class
of theories in which all the mass-scales (at least 
during the epoch of interest) are set by $\phi$ alone. In such a case
$M(\phi) \propto \phi,\, \Gamma(\phi) \propto \phi$, and density perturbations
via our mechanism acquire an especially simple form, practically 
independent of either
the coupling constants or the field content. Interestingly, the Standard 
Model,
as well as its minimal supersymmetric extension, in which $\phi$
is identified with a flat direction field,  
fall within this category.

Before proceeding we have to stress that there can be corrections to
the masses and decay rates from other existing scales in the theory. 
For instance, non-perturbative gravity effects can
contribute into the decay rates of the Standard Model particles. 
If such corrections are there in the first place, their
relative value will depend on many factors, such as existence of
discrete gauge symmetries, masses of decaying particles etc. It is 
not hard within 
a concrete model to make this corrections sub-dominant. This
is beyond the scope of the present work, in which we shall deal only with
the simplest minimal case. 

We shall now apply our mechanism to Standard Model, and to its minimal
supersymmetric extension.  We assume that {\it (1)} during inflation
at least some of the MSSM flat direction fields (call them $\phi$)
have masses $\ll H$; and {\it (2)} after the inflaton decay, the
Universe is left with a thermal gas of SM particles (and their
superpartners) of uniform energy density.  Under these assumptions we
will demonstrate that the amount of the density perturbations
generated through the cooling process, {\it irrespective} of the
coupling constants or the nature of flat direction, is given by
\begin{equation}
{\delta \rho \over\rho} \, =  \, 
{2\over 3}{ \delta \phi \over \phi },
\label{deltasm}
\end{equation}
provided some of the particles dominate the Universe for a short period. 

 To show this, let us first ignore superpartners and only consider
Standard Model particles. Let $\phi$ be a flat direction corresponding
to the Standard Model Higgs field.  That is, we shall assume that the
value of the Higgs mass during inflation satisfies our requirements,
and that the Universe is reheated by producing some of the standard
model particles. When the Universe cools down, the heavy particles
decouple and decay into the lighter ones. The heavy states that have
small annihilation and decay rates can dominate the Universe.  Such
can only be fermions (quarks and leptons), since the gauge bosons
decay through the order one gauge couplings and decay before
domination.

For a given species to dominate it must have frozen out but still have
not had enough time to decay. Thus both its annihilation and decay
rates must be less than $H$.  This situation can always be achieved
for some species for the appropriate initial value of $\phi$.  Assume
for definiteness that $\phi \sim M_P$, and, of course, we shall assume
that $T_R \ll M_P$, but big enough that some of the unstable fermions
are produced.  We shall now discuss under what conditions some of the
species can undergo a brief interval of domination. It is useful to
consider electrically charged fermions and the neutral ones
separately.

Charged fermions can annihilate into photons so their freeze-out
abundance will be much smaller than $M^3$ unless their mass is of
order $ M \sim \alpha_{EM}^2 M_{Pl}$.  Their mass however has to be
smaller than the reheating temperature at the end of inflation for
them to be in thermal equilibrium to start with, so $M <
T_{rh}\lesssim \sqrt{H_{inf} M_{PL}} \lesssim 10^{-2} M_{PL}$.  This
case is only marginally possible so we will concentrate on uncharged
fermions, ie. neutrinos.


The story with neutrinos is very different, as we shall now
discuss. In order to understand how neutrinos can dominate in very
early Universe, we shall specify the origin of their mass.  For
definiteness, we shall assume the standard ``See-Saw''
mechanism\cite{seesaw}, in which the neutrino masses are generated by
mixing to the heavy gauge-singlet fermions (right-handed neutrinos).
The relevant terms in the Lagrangian are
\begin{equation}
\label{neutrino}
  \lambda_{\nu}\, \phi\, \bar{\nu}_L \nu_R \, + \, M_R\,\nu_R \nu_R \, + \, ...
\end{equation}
where $M_R$ is the Majorana mass of the gauge-singlet fermion
$\lambda$ is an Yukawa coupling constant, and generation indicies are
suppressed.  

In today's Universe
\begin{equation}
\lambda_{\nu}\phi \ll M_R, 
\label{neutrino1}
\end{equation}
and heavy neutrinos can be integrated out. As a result of this integration the light neutrinos
acquire small Majorana masses given by
\begin{equation}
M_{\nu} \sim {(\lambda_{\nu}\phi)^2 \over M_R} 
\label{Majorana}
\end{equation}
However, in the epoch of our interest $\phi \sim M_{Pl}$, and the condition (\ref{neutrino1}) can be violated.
Depending whether this is the case, the light neutrino will either continue to be a Majorana particle
with mass (\ref{Majorana}), or will effectively become a Dirac particle of mass
\begin{equation}
M_{\nu} \sim \lambda_{\nu}\phi 
\label{Dirac}
\end{equation}
Neutrino annihilation rate into the fermions happens through the exchange of heavy $Z,W$ bosons, with masses
$\sim M_{Pl}$, and the annihilation into the Higgs particles 
is suppressed by $\lambda_{\nu}^4/M_R^2$.
So it is safe to assume that their abundance  froze out early on.

The decay rate of neutrinos requires some attention. In the early Universe some of the neutrinos that are stable today,
could have been unstable. For instance, since the electron and quark
masses scale linearly with $\phi$ as opposed to neutrino masses that
scale quadratically, the electron neutrino $\nu_e$ can become heavier
and decay into the electron and the quark-anti-quark pair
\begin{equation}
\label{nu}
\nu_e \rightarrow \, e^{-} + u + \bar{d} 
\end{equation}

In general, the Standard Model fermions decay through the exchange of $W$-bosons and the decay rate is:
\begin{equation}
\label{fermi}
\Gamma =  f M^5/\phi^4 
\end{equation}
Where $f$ is a small constant that depends on mixing angles, and is different for different fermions. The crucial point is that because $M$ is set by  $\phi$,  $\Gamma$ is linear in
$\phi$ (for neutrino this requires $\lambda_{\nu}\phi > M_R$)

\begin{equation}
\label{fermilinear}
\Gamma = f \lambda^5 \phi
\end{equation} 
As a result of this,  irrespective which fermion happens to dominate, the imprint of density perturbation is {\it universally} given by (\ref{deltasm}).  It is easy to check that for 
$\phi \sim M_{Pl} \sim 10^4 H$, the domination condition is always satisfied by at least some SM fermions.
For large values of $\phi$ and $H$, our mechanism can operate several times and each time
imprint the density perturbations. 

Interestingly, going to MSSM does not modify this general result,
irrespective which flat direction develops large VEV during inflation. This
can be understood in the following way. Let $\phi$ be some MSSM flat
direction. $\phi$ breaks gauge symmetry and gives masses $\propto
\phi$ to some gauge fields and fermions (and their
superpartners). Since, for the needed magnitude of perturbations the
flat direction VEV must be $>> H$, the supersymmetry breaking effects
can be ignored. So we can neglect the mass splittings between the
superpartners.  Thus, irrespective of the particular flat direction,
the $\phi$-dependence of $\Gamma$ will be the same.  This is obvious
since the masses of particles are set by $\phi$, and heavy particles
decay into the light ones through the exchange of the massive gauge
bosons. \footnote{The massless exchange cannot lead to a particle
decay, due to the fact that the massless gauge bosons couple through
unbroken generators, which commute with the Hamiltonian.} This leads us to a
conclusion that the density perturbations generated via our
mechanism is independent of the detailed structure of the couplings as
well as the nature of dominating flat directions, and is always given
by equation (\ref{deltasm}).

\section{Non-Gaussianities}

The WMAP satellite has provided bounds for the degree of
non-Gaussianity of the primordial cosmological perturbations. Assuming
that the super-horizon gravitational potential fluctuations (during
the matter-dominated era) are of the form
\begin{equation}
\Phi = g +f_{nl}~g^2,
\end{equation}
where $g$ is Gaussian, the WMAP results are $-58<f_{nl}<134$ at 95\%
confidence \cite{komatsu}. The Sloan Digital Sky Survey should provide
similar accuracy \cite{roman}.
 
The standard one-field slow-roll inflation predicts the degree of
non-Gaussianity which corresponds to $f_{nl}\sim$ tilt of the
perturbation spectrum \cite{maldacena}. The power spectrum measured by
WMAP is consistent with scale invariant, however when WMAP data is
combined with other probes of large scale structure small tilts maybe
preferred \cite{spergel}. It is fair to say that the largest tilts
still allowed by the data are of order $|n-1| \lesssim 0.05$. Thus the
current theoretical expectation is that primordial non-Gaussianity is
unobservably small. We will show that in our model non-Gaussianities
are much larger.

We will consider non-Gaussianities assuming that the fluctuations of
the light field $\phi$ which is responsible for fluctuations of the
coupling constants and/or masses are purely Gaussian. The case of
non-Gaussianities of $\phi$ will be discussed elsewhere.

To talk about non-Gaussianities, one needs to define a gauge-invariant quantity to quadratic order. We will follow \cite{maldacena,bst,sb}, and use a gauge invariant variable $\zeta$ that remains constant outside the horizon. It is defined as follows: $e^{\zeta }$ is proportional to the local scale factor measured on uniform local Hubble parameter hypersurfaces. To linear order, $\zeta$ is proportional to the gravitational potential $\Phi$. During matter domination $\zeta =-(5/3)\Phi$, during radiation domination $\zeta =-(3/2)\Phi$ \cite{mukh}.

In our original scenario \cite{dgz}, the energy density fluctuations on hypersurfaces of constant scale factor $a$ are
\begin{equation}
\rho \propto \Gamma ^{-2/3}.
\end{equation}
Since $\rho \propto a^{-4}$, this gives 
\begin{equation}
\zeta=-(1/6)\log \Gamma .
\end{equation}

The inflaton reheating rate $\Gamma =\lambda ^2m_\phi$, where $m_\phi$
is the inflaton mass at its minimum, and $\lambda$ is the coupling
constant of the inflaton decay. We have to consider two case. First
assume that fluctuations of $\lambda$ are negligible, and
$m_\phi\propto \phi$ dominates the fluctuations of $\Gamma$. Then, up
to second order,
\begin{equation}
\zeta=-{1\over 6}\left( \delta _\phi - {1\over 2}\delta _\phi ^2\right),
\end{equation}
where $\delta _\phi\equiv (\phi /<\phi >-1)$ is assumed to be Gaussian.
The definition of $f_{nl}$ used by \cite{komsper} corresponds to \cite{maldacena}
\begin{equation}
\zeta=g-(3/5)f_{nl}g^2,
\end{equation}
And gives 
\begin{equation}
f_{nl}=-5
\end{equation}
for this scenario. If, on the other hand, $\lambda \propto \phi$ dominates the fluctuations of the reheating rate, one obtains a smaller non-Gaussianity, 
\begin{equation}
f_{nl}=-5/2.
\end{equation}
This shows the general trend: less efficient mechanisms of translation of the $\phi$-fluctuation into the metric fluctuation give greater non-Gaussianities (assuming the inefficient mechanism is still the dominant one).  Our assumptions that $m\propto \phi$ and $\lambda \propto \phi$ can also be generalized to include non-linear terms, leading  different values of $f_{nl}$.

The new mass-domination scenario described in \S II is of particular
interest because of its high universality, based on $\phi$ being the
only mass scale of the theory. From $\rho \propto \phi ^{2/3}$, we get
a unique answer independent of the Yukawa-couplings:
\begin{equation}
f_{nl}=5.
\end{equation}

\section{Conclusions}

We proposed a new scenario for generation of metric perturbations
after inflation, whereby superhorizon fluctuations of the light fields
generated during inflation are translated into metric fluctuations
during reheating, mass domination, or freeze-out.

We calculated non-Gaussianities for this model. Our result is
$f_{nl}=5$ for the mass domination mechanism. Other scenarios give
different values, and non-Gaussianities of the light fields $\phi$
also lead to partial compensation or enhancement of the intrinsic
non-Gaussianities of our mechanism, but an observable value
$|f_{nl}|\sim$few is a natural lower bound.

\bigskip
{\bf Acknowledgments}

We thank J. Maldacena, V. F.  Mukhanov, P.J.E.Peebles, R. Scoccimarro for useful discussions. 

The work of G.D., A.G., and M.Z. is supported by the David and Lucile  Packard Foundation. The work of G.D. is also supported by the Alfred P. Sloan foundation and by NSF PHY-0070787.

\end{document}